\documentclass[reqno,12pt,a4paper]{amsart}

\usepackage{mathrsfs}
\usepackage{amssymb}
\usepackage{amsfonts}
\usepackage{latexsym}
\usepackage{amsthm}

\usepackage{graphicx}
\def\lb{\label}

\newcommand{\er}[1]{\textrm{(\ref{#1})}}

\begin{document}


\renewcommand{\theequation}{\arabic{section}.\arabic{equation}}
\theoremstyle{plain}
\newtheorem{theorem}{\bf Theorem}[section]
\newtheorem{lemma}[theorem]{\bf Lemma}
\newtheorem{corollary}[theorem]{\bf Corollary}
\newtheorem{proposition}[theorem]{\bf Proposition}
\newtheorem{definition}[theorem]{\bf Definition}
\newtheorem{remark}[theorem]{\it Remark}

\def\a{\alpha}  \def\cA{{\mathcal A}}     \def\bA{{\bf A}}  \def\mA{{\mathscr A}}
\def\b{\beta}   \def\cB{{\mathcal B}}     \def\bB{{\bf B}}  \def\mB{{\mathscr B}}
\def\g{\gamma}  \def\cC{{\mathcal C}}     \def\bC{{\bf C}}  \def\mC{{\mathscr C}}
\def\G{\Gamma}  \def\cD{{\mathcal D}}     \def\bD{{\bf D}}  \def\mD{{\mathscr D}}
\def\d{\delta}  \def\cE{{\mathcal E}}     \def\bE{{\bf E}}  \def\mE{{\mathscr E}}
\def\D{\Delta}  \def\cF{{\mathcal F}}     \def\bF{{\bf F}}  \def\mF{{\mathscr F}}
\def\c{\chi}    \def\cG{{\mathcal G}}     \def\bG{{\bf G}}  \def\mG{{\mathscr G}}
\def\z{\zeta}   \def\cH{{\mathcal H}}     \def\bH{{\bf H}}  \def\mH{{\mathscr H}}
\def\e{\eta}    \def\cI{{\mathcal I}}     \def\bI{{\bf I}}  \def\mI{{\mathscr I}}
\def\p{\psi}    \def\cJ{{\mathcal J}}     \def\bJ{{\bf J}}  \def\mJ{{\mathscr J}}
\def\vT{\Theta} \def\cK{{\mathcal K}}     \def\bK{{\bf K}}  \def\mK{{\mathscr K}}
\def\k{\kappa}  \def\cL{{\mathcal L}}     \def\bL{{\bf L}}  \def\mL{{\mathscr L}}
\def\l{\lambda} \def\cM{{\mathcal M}}     \def\bM{{\bf M}}  \def\mM{{\mathscr M}}
\def\L{\Lambda} \def\cN{{\mathcal N}}     \def\bN{{\bf N}}  \def\mN{{\mathscr N}}
\def\m{\mu}     \def\cO{{\mathcal O}}     \def\bO{{\bf O}}  \def\mO{{\mathscr O}}
\def\n{\nu}     \def\cP{{\mathcal P}}     \def\bP{{\bf P}}  \def\mP{{\mathscr P}}
\def\r{\rho}    \def\cQ{{\mathcal Q}}     \def\bQ{{\bf Q}}  \def\mQ{{\mathscr Q}}
\def\s{\sigma}  \def\cR{{\mathcal R}}     \def\bR{{\bf R}}  \def\mR{{\mathscr R}}
                \def\cS{{\mathcal S}}     \def\bS{{\bf S}}  \def\mS{{\mathscr S}}
\def\t{\tau}    \def\cT{{\mathcal T}}     \def\bT{{\bf T}}  \def\mT{{\mathscr T}}
\def\f{\phi}    \def\cU{{\mathcal U}}     \def\bU{{\bf U}}  \def\mU{{\mathscr U}}
\def\F{\Phi}    \def\cV{{\mathcal V}}     \def\bV{{\bf V}}  \def\mV{{\mathscr V}}
\def\P{\Psi}    \def\cW{{\mathcal W}}     \def\bW{{\bf W}}  \def\mW{{\mathscr W}}
\def\o{\omega}  \def\cX{{\mathcal X}}     \def\bX{{\bf X}}  \def\mX{{\mathscr X}}
\def\x{\xi}     \def\cY{{\mathcal Y}}     \def\bY{{\bf Y}}  \def\mY{{\mathscr Y}}
\def\X{\Xi}     \def\cZ{{\mathcal Z}}     \def\bZ{{\bf Z}}  \def\mZ{{\mathscr Z}}
\def\O{\Omega}

\newcommand{\gA}{\mathfrak{A}}
\newcommand{\gB}{\mathfrak{B}}
\newcommand{\gC}{\mathfrak{C}}
\newcommand{\gD}{\mathfrak{D}}
\newcommand{\gE}{\mathfrak{E}}
\newcommand{\gF}{\mathfrak{F}}
\newcommand{\gG}{\mathfrak{G}}
\newcommand{\gH}{\mathfrak{H}}
\newcommand{\gI}{\mathfrak{I}}
\newcommand{\gJ}{\mathfrak{J}}
\newcommand{\gK}{\mathfrak{K}}
\newcommand{\gL}{\mathfrak{L}}
\newcommand{\gM}{\mathfrak{M}}
\newcommand{\gN}{\mathfrak{N}}
\newcommand{\gO}{\mathfrak{O}}
\newcommand{\gP}{\mathfrak{P}}
\newcommand{\gQ}{\mathfrak{Q}}
\newcommand{\gR}{\mathfrak{R}}
\newcommand{\gS}{\mathfrak{S}}
\newcommand{\gT}{\mathfrak{T}}
\newcommand{\gU}{\mathfrak{U}}
\newcommand{\gV}{\mathfrak{V}}
\newcommand{\gW}{\mathfrak{W}}
\newcommand{\gX}{\mathfrak{X}}
\newcommand{\gY}{\mathfrak{Y}}
\newcommand{\gZ}{\mathfrak{Z}}

\def\ve{\varepsilon} \def\vt{\vartheta} \def\vp{\varphi}  \def\vk{\varkappa}
\def\vr{\varrho}

\def\Z{{\mathbb Z}} \def\R{{\mathbb R}} \def\C{{\mathbb C}}  \def\K{{\mathbb K}}
\def\T{{\mathbb T}} \def\N{{\mathbb N}} \def\dD{{\mathbb D}} \def\S{{\mathbb S}}
\def\B{{\mathbb B}}


\def\la{\leftarrow}              \def\ra{\rightarrow}     \def\Ra{\Rightarrow}
\def\ua{\uparrow}                \def\da{\downarrow}
\def\lra{\leftrightarrow}        \def\Lra{\Leftrightarrow}
\newcommand{\abs}[1]{\lvert#1\rvert}
\newcommand{\br}[1]{\left(#1\right)}

\def\lan{\langle} \def\ran{\rangle}


\def\lt{\biggl}                  \def\rt{\biggr}
\def\ol{\overline}               \def\wt{\widetilde}
\def\no{\noindent}


\let\ge\geqslant                 \let\le\leqslant
\def\lan{\langle}                \def\ran{\rangle}
\def\/{\over}                    \def\iy{\infty}
\def\sm{\setminus}               \def\es{\emptyset}
\def\ss{\subset}                 \def\ts{\times}
\def\pa{\partial}                \def\os{\oplus}
\def\om{\ominus}                 \def\ev{\equiv}
\def\iint{\int\!\!\!\int}        \def\iintt{\mathop{\int\!\!\int\!\!\dots\!\!\int}\limits}
\def\el2{\ell^{\,2}}             \def\1{1\!\!1}
\def\sh{\sharp}
\def\wh{\widehat}
\def\bs{\backslash}
\def\na{\nabla}

\def\sh{\mathop{\mathrm{sh}}\nolimits}
\def\all{\mathop{\mathrm{all}}\nolimits}
\def\Area{\mathop{\mathrm{Area}}\nolimits}
\def\arg{\mathop{\mathrm{arg}}\nolimits}
\def\const{\mathop{\mathrm{const}}\nolimits}
\def\det{\mathop{\mathrm{det}}\nolimits}
\def\diag{\mathop{\mathrm{diag}}\nolimits}
\def\diam{\mathop{\mathrm{diam}}\nolimits}
\def\dim{\mathop{\mathrm{dim}}\nolimits}
\def\dist{\mathop{\mathrm{dist}}\nolimits}
\def\Im{\mathop{\mathrm{Im}}\nolimits}
\def\Iso{\mathop{\mathrm{Iso}}\nolimits}
\def\Ker{\mathop{\mathrm{Ker}}\nolimits}
\def\Lip{\mathop{\mathrm{Lip}}\nolimits}
\def\rank{\mathop{\mathrm{rank}}\limits}
\def\Ran{\mathop{\mathrm{Ran}}\nolimits}
\def\Re{\mathop{\mathrm{Re}}\nolimits}
\def\Res{\mathop{\mathrm{Res}}\nolimits}
\def\res{\mathop{\mathrm{res}}\limits}
\def\sign{\mathop{\mathrm{sign}}\nolimits}
\def\span{\mathop{\mathrm{span}}\nolimits}
\def\supp{\mathop{\mathrm{supp}}\nolimits}
\def\Tr{\mathop{\mathrm{Tr}}\nolimits}
\def\BBox{\hspace{1mm}\vrule height6pt width5.5pt depth0pt \hspace{6pt}}
\def\where{\mathop{\mathrm{where}}\nolimits}
\def\as{\mathop{\mathrm{as}}\nolimits}


\newcommand\nh[2]{\widehat{#1}\vphantom{#1}^{(#2)}}
\def\dia{\diamond}

\def\Oplus{\bigoplus\nolimits}



\def\qqq{\qquad}
\def\qq{\quad}
\let\ge\geqslant
\let\le\leqslant
\let\geq\geqslant
\let\leq\leqslant
\newcommand{\ca}{\begin{cases}}
\newcommand{\ac}{\end{cases}}
\newcommand{\ma}{\begin{pmatrix}}
\newcommand{\am}{\end{pmatrix}}
\renewcommand{\[}{\begin{equation}}
\renewcommand{\]}{\end{equation}}
\def\eq{\begin{equation}}
\def\qe{\end{equation}}
\def\[{\begin{equation}}
\def\bu{\bullet}
\newcommand{\fr}{\frac}
\newcommand{\tf}{\tfrac}

\title[{Asymptotics of resonances for 1d Stark operators}]
{Asymptotics of resonances for 1d Stark  operators}

\date{\today}
\author[Evgeny Korotyaev]{Evgeny L. Korotyaev}
\address{Saint-Petersburg State University, Universitetskaya nab.
7/9, St. Petersburg, 199034, Russia, \ korotyaev@gmail.com, \
e.korotyaev@spbu.ru}


\begin{abstract}
\no We consider the Stark operator perturbed by a compactly
supported potentials on the real line. We
determine forbidden domain for resonances, asymptotics of resonances at high energy
 and asymptotics of the resonance counting function for large radius.
\end{abstract}

\maketitle




\section {Introduction and main results}
\setcounter{equation}{0}

\subsection{Introduction}
We consider the operator $H=H_0+V$ acting on
$L^2(\R)$, where the  unperturbed  operator $H_0=-{d^2\/dx^2}+  x$ is   the  Stark
operator.
Here  $x$ is an external electric field and the potential $V=V(x),
x\in \R$ is real and satisfies
\[
\label{1.2}
V\in L_{real}^2(\R),\qqq \supp V\ss [0,\g]\qq {\rm for\ some\ }\g>0.
\]
The operators $H_0$ and $H$ are self-adjoint on the
same domain since the operator $V(H_0-i)^{-1}$ is compact, see e.g. \cite{RS76,J85,L93, K17}.
The spectrum of both $H_0$ and $H$ is purely absolutely continuous
and covers the real line $\R$ (see \cite{RS76,J85,L93}).
It is well known that the wave operators $W_\pm$ for
the pair $H_0, H$ given by
$$
W_\pm=s-\lim e^{itH}e^{-itH_0} \qqq \as \qqq t\to \pm\iy,
$$
exist and  are unitary (even for more large class of potentials than considered here, see \cite{RS76,J85,L93}).  Thus the scattering operator $S=W_+^*W_-$ is unitary.
The operators $H_0$
and $S$ commute and then are simultaneously diagonalizable:
\[
\lb{DLH0}
L^2(\R)=\int_\R^\oplus \mH_\l d\l,\qqq H_0=\int_\R^\oplus\l I_\l d\l,\qqq
S=\int_\R^\oplus S(\l)d\l;
\]
here $I_\l$ is the identity in the fiber space $\mH_\l=\C$ and $S(\l)$ is
the scattering matrix (which is a  scalar function of $\l\in \R$ in our case)
 for the pair $H_0, H$.
  The function $S(\l)$ is continuous in $\l\in \R$ and satisfies
$S(\l)=1+o(1)$ as $\l\to \pm\iy$. The function $S(\l)$ has an analytic extension into the upper half-plane $\C_+=\{\l\in\C: \Im \l>0\}$ and a meromorphic extension into the lower half-plane $\C_-=\{\l\in\C: \Im \l<0\}$, see e.g. \cite{K17}. By definition, a zero $\l_o\in \C_+$ (or a pole $\l_o\in \C_-$) of $S$ is called a resonance.
The multiplicity of the resonance is the multiplicity of
  the corresponding zero (or a pole)  of $S$.


\medskip

\no {\bf Condition C}. {\it The potential  $V$  satisfies \er{1.2} and a following asymptotics
 \[
\label{CV2}
\begin{aligned}
\int_0^\g e^{i2xk} V(x)dx=\fr{C_p}{(-i2k)^p}+\fr{O(1)}{k^\n},\\
\end{aligned}
\]
as $k\in \ol\C_+,\ |k|\to \iy$
 uniformly in $\arg k\in [0,\pi]$, where $p\in ({1\/2},1)$,  $ p<\n$ and $(-ik)^p=e^{-ip{\pi\/2}}k^p$.}

{\bf Remark.} 1) Here and below for $\a>0, \l \in \ol\C_+$ we define
\[
\lb{GRi}
\begin{aligned}
\l^\a=|\l|^a e^{i\a\arg \l},\qqq \arg \l\in [0,\pi]
 \qq
  \log \l=\log |k|+i\arg \l.
\end{aligned}
\]
2) Let $V(x)=\fr{C_*}{x^{1-p}}+V_1(x), x\in [0,\g]$ for some $p\in ({1\/2},1)$, where
$V, V_1, V_1'$ satisfy \er{1.2}. Then $V$ satisfies \er{CV2} with $C_p=C_*\G(p)$, see Lemma \ref{TaV}.


\subsection{Main results}
Resonances for the Stark operator perturbed by a compactly
supported potential (of a certain class)  on the real line were considered in \cite{K17}.
The following results were proved:\\
$\bu$  upper and lower bounds on the number of resonances in
complex discs with large radii,\\
$\bu$  the trace formula  in
terms of resonances only, \\
$\bu$  it was shown that all resonances determine the potential uniquely.

Our main goal is to obtain global properties of the resonances of $H$.
We describe the forbidden domain for resonances.

\begin{theorem}
\lb{T1}
Define the function
$
\x(\l)={e^{-i{4\/3} \l^{3\/2}}\/2\sqrt \l}, \l\in \ol\C_+$, where  $\sqrt \l\in \ol\C_+$
 and let $|\l|\to \iy$.

i) Let $V$ satisfy \er{1.2} and  $\arg \l\in [\fr{2\pi}{3}, {\pi}]$ . Then   $S(\l)$  satisfies
\[
\lb{aD1}
\begin{aligned}
S(\l)=1+\x(\l)O(1),
\end{aligned}
\]
and there are no resonances in the set $\{\vp\in [\fr{2\pi}{3}, {\pi}],|\l|\ge \vr \}$
for some $\vr>0$ large enough.

ii) Let $V$ satisfy \er{CV2} and let
 $\arg \l\in [\ve, \fr{2\pi}{3}-\ve]$ for any $\ve>0$. Then
$S(\l)$  satisfies
\[
\lb{aD2}
\begin{aligned}
S(\l)=\x(\l)\fr{C_p+o(1)}{(-i2k)^p},
\end{aligned}
\]
and there are no resonances in the set $\{\vp\in[\ve, \fr{2\pi}{3}-\ve],|\l|\ge \vr \}$
for some $\vr>0$ large enough.

\end{theorem}


{\bf Remark.} 1) Thus we have  $S(\l)\to 1$ in \er{aD1} and
$|S(\l)|\to \iy$ in \er{aD2}.

2) Consider the Schr\"odinger operator $hy=-y''+Vy, y(0)=0$ with a
compactly supported potential $V$ on the half-line. The resolvent
$(h-\l)^{-1}$ has a meromorphic  extension from the first sheet
$\L_1=\C\sm [0,\iy)$ of the two sheeted simple Riemann surface of
the function $\sqrt \l$ on the second sheet $\L_2=\C\sm [0,\iy)$.
Each pole defines the resonance  and the set of resonances is
symmetric with respect to the real line. We consider resonances in
$\C_+\ss \L_2$. Here there are infinitely many resonances (see
\cite{Z87}) and multiplicity of a resonance can be any number (see
\cite{K04}). Denote by $\gN(r)$ the number of resonances having
modulus $\leq r$, each zero being counted according to its
multiplicity. Zworski's result \cite{Z87} gives

\[
\lb{za}  \gN(r)= {2\/ \pi}r^{1\/2}(\g+o(1))\ \ \as \qq r\to\iy.
\]

 Moreover, each  resonance $\l_o$ satisfies
$|\l_o|\leq C_0^2e^{4|\Im \sqrt{\l_o}|}$, where
$C_0=\|q\|_1e^{\|q\|_1}$ and  $\|q\|_1=\int_{\R_+} |q(x)|dx$ (see
\cite{K04}). This  gives the forbidden (so-called logarithmic)
domain $\{\l\in \C_+: |\l|> C_0^2e^{4|\Im \sqrt{\l}|}\}$ for the
resonances.  Note that due to \er{aD1} the forbidden domain for $H$
has the form $\{\l\in \C_+: |\l|>\r, \arg \l\in [{2\pi\/3},\pi]\}$
for some $\r>0$, see Fig. \ref{Fig1}.


\setlength{\unitlength}{1.0mm}
\begin{figure}[h]
\centering
\unitlength 0.9mm 
\linethickness{0.4pt}
\ifx\plotpoint\undefined\newsavebox{\plotpoint}\fi 
\begin{picture}(160,80)(0,0)
\hspace{-10mm} \put(40,40){\vector(1,0){110.00}}
\put(80,2){\vector(0,1){76.00}} \put(80,40){\line(-2,3){24.0}}
\put(80,40){\line(-2,-3){24.0}}

\bezier{600}(60,40)(61,59)(80,60) \bezier{600}(80,60)(99,59)(100,40)
\bezier{600}(60,40)(61,21)(80,20) \bezier{600}(80,20)(99,21)(100,40)

\bezier{25}(58,76)(68,65)(75,59.5)
\bezier{60}(98.5,48)(120,43)(140,42)

\bezier{25}(58,4)(68,15)(75,20.5)
\bezier{60}(98.5,32)(120,37)(140,38)


\bezier{600}(70,40)(70,45)(74,49.0)
\bezier{600}(70,40)(70,35)(74,31.0)

\put(67.5,45){$\scriptstyle{\pi\/3}$}
\put(67.5,33.5){$\scriptstyle{\pi\/3}$}

\put(63,46){$\scriptstyle\ts$} \put(63,32.5){$\scriptstyle\ts$}

\put(71.5,53){$\scriptstyle\ts$} \put(76,55){$\scriptstyle\ts$}
\put(71.5,25.5){$\scriptstyle\ts$} \put(76,23.5){$\scriptstyle\ts$}

\put(85,45){$\scriptstyle\ts$} \put(83,54){$\scriptstyle\ts$}
\put(90,50){$\scriptstyle\ts$}

\put(85,33.5){$\scriptstyle\ts$} \put(83,24.5){$\scriptstyle\ts$}
\put(90,28.5){$\scriptstyle\ts$}

\put(100,69){$\C_+$ -- second sheet} \put(105,63){zeros of $S(\l)$}

\put(100,13){$\C_-$ -- second sheet} \put(105,7){poles of $S(\l)$}

\put(59,75){\circle*{1.0}} \put(62,71.7){\circle*{1.0}}
\put(65,68.5){\circle*{1.0}} \put(68.5,65.3){\circle*{1.0}}
\put(72.0,62.1){\circle*{1.0}}

\put(59,5){\circle*{1.0}} \put(62,8.3){\circle*{1.0}}
\put(65,11.5){\circle*{1.0}} \put(68.5,14.7){\circle*{1.0}}
\put(72.0,17.9){\circle*{1.0}}

\put(105,46.6){\circle*{1.0}} \put(111,45.4){\circle*{1.0}}
\put(117,44.4){\circle*{1.0}} \put(123,43.6){\circle*{1.0}}
\put(129,42.8){\circle*{1.0}} \put(135,42.2){\circle*{1.0}}

\put(105,33.4){\circle*{1.0}} \put(111,34.6){\circle*{1.0}}
\put(117,35.6){\circle*{1.0}} \put(123,36.4){\circle*{1.0}}
\put(129,37.2){\circle*{1.0}} \put(135,37.8){\circle*{1.0}}

\end{picture}
\caption{\footnotesize } \label{Fig1} Resonances on the non-physical
sheet $\C_+$ (on $\C_-$) are zeros (poles ) of $S(\l)$.
\end{figure}

\subsection{Asymptotics of resonances} Define numbers
\[
\lb{dxn}
z_*^+={\pi\/2}(p+2)+i\log {3^b\/C_p},\qq
                                  z_*^-=z_*^+-b\pi,\qq b={p+1\/3},\qquad
\r_r=\pi(2r-1)+\Re z_*^+-{b\pi\/2}.
\]

\begin{theorem}
\lb{T2} Let  $V$ satisfy Condition {\rm C} and let $b,z_*,\r_r$ be given by \er{dxn}
for some integer $r>1$   large enough.
 Then the function $S(\l)$ in the domain $\cD_r=\C_+\sm \{|\l|<\r_r^{2\/3}\}$ have only simple zeros $\l_n^\pm, n\ge r$ labeled  by $|\l_r^\pm|<|\l_{r+1}^\pm|<...$ with asymptotics
\[
\lb{arn1}
\begin{aligned}
\l_n^\pm=\rt(\fr{\pm 3\pi n}2\rt)^{2\/3}\rt(1\pm \fr{(ib\log |2\pi n|+z_*^\pm)}{3\pi n}+\fr{O(1)}{n^{1+{s\/3}}}\rt)  \rt)\qq \as \qq n\to \iy
\end{aligned}
\]
for any $s<\min\{\n,1\}-p$, where $(1)^{2\/3}=1$ and $(-1)^{2\/3}=e^{i{2\pi\/3}}$.
\end{theorem}

{\bf Remarks.} Note that $\Im \l_n^+=c\fr{\log n}{n^{1\/3}}+...$ as $n\to \iy$.
Thus the sequence of the resonances $\l_n^+\in \C_+$ on the second sheet is more and more close to the real line.
At the same time we have $\| Y(\l)\|\to 0$ as $|\l|\to \iy$ on the first sheet $\l\in \ol\C_-$, see \er{e.2}.  It means that
perturbed resolvent has the residues at the simple resonances $\l_n^+$, which go zero very fast  at large $n\to \iy$.

Denote by $\cN(r)$  the number of zeros in $\C_+$ (resonances of $H$)  of
$S$ having modulus $\leq r$ and counted according to multiplicity.

\begin{corollary}
\lb{T3} Let  $V$ satisfy Condition {\rm C}.
Then the counting function $\cN(r)$ satisfies
\[
\lb{Nr}
\cN(r)=\fr{4r^{3\/2}}{3\pi}
(1+ o(1))\qqq\as\qq r\to\iy.
\]

\end{corollary}

{\bf Remarks.} 1)   In the Zworski asymptotics \er{za}  the first terms depends on the
 diameter of support of a potential. In the Stark case \er{Nr} the first  term  does not depends on the potential.

2) Roughly speaking, the number of resonances of the perturbed Stark operator $H$ on the real line corresponds
to one  for the Schr\"odinger operator on $\R^3$.

\subsection{Brief overview}
A lot of papers are devoted to resonances of the one-dimensional
Schr\"odinger operator, see Froese \cite{F97}, Hitrik \cite{H99}, Korotyaev \cite{K04},
Simon \cite{S00}, Zworski \cite{Z87} and references given there.
 Inverse problems (characterization, recovering,
uniqueness) in terms of resonances were solved by Korotyaev for a
Schr\"odinger operator with a compactly supported potential on the
real line \cite{K05} and the half-line \cite{K04}, see also Zworski
\cite{Z02}, Brown-Knowles-Weikard \cite{BKW03} concerning the
uniqueness. The resonances for  one-dimensional operators
$-{d^2\/dx^2}+V_\pi+V$, where $V_\pi$ is periodic and $V$ is a
compactly supported potential were considered  by Firsova
\cite{F84}, Korotyaev \cite{K11}, Korotyaev-Schmidt \cite{KS12}.
Christiansen \cite{C06} considered resonances for steplike potentials.
 Lieb-Thirring type inequality for the resonances was determined in
 \cite{K16}. The "local resonance" stability problems were considered in
\cite{Ko04}, \cite{MSW10}.

Next, we mention some results  for  one-dimensional perturbed Stark
operators: \\
$\bu$ the scattering theory was considered by
Rejto-Sinha \cite{RS76}, Jensen \cite{J85},   Liu  \cite{L93};\\
$\bu$ the inverse scattering problem are studied by
Calogero-Degasperis \cite{CD78},
Kachalov-Kurylev \cite{KK91}, Kristensson \cite{K86}, Lin-Qian-Zhang
\cite{LQZ89};\\
$\bu$ there are a lot of results about the resonances, where the dilation
analyticity techniques are used, see e.g., Herbst \cite{H79}, Jensen \cite{J89}
 and references therein. Note that compactly
supported potentials are not treated in these papers.\\
$\bu$ There are interesting results about resonances for
one-dimensional Stark-Wannier operators $-{d^2\/dx^2}+\ve x+V_\pi$,
where the constant $\ve>0$ is the electric field strength and $V_\pi$ is
the real periodic potential: Agler-Froese \cite{AF85},
 Herbst-Howland \cite{HH79},  Jensen
\cite{J86}.

Finally we note the  that  Herbst and Mavi \cite{HM16} considered
resonances of  the Stark operator perturbed by delta-potentials.

\subsection{Plan of the paper}
In Section 2 we recall well known results on   basic estimates for
the Stark operator  in a form useful for our approach. In  Section 3
we prove the main results.  The Appendix contains technical
estimates needed  to obtain sharp asymptotics \eqref{arn1}.

\section {Properties of S-matrix}
\setcounter{equation}{0}

\medskip

\subsection{The well-known facts.}
 We denote by $C$
various possibly different constants whose values are immaterial in
our constructions.
 We introduce resolvents $R(\l)=(H-\l)^{-1}$ and
 $R_0(\l)=(H_0-\l)^{-1}$ and operators  $Y, Y_0$ by
\[
\lb{Y0}
\begin{aligned}
Y(\l) = |V|^{1\/2}R(\l)V^{1\/2},\qqq Y_0(\l) =
|V|^{1\/2}R_0(\l)V^{1\/2},\qq \l\in \C_\pm,\\
V=|V|^{1\/2}V^{1\/2},\qqq V^{1\/2}=|V|^{1\/2} \sign V.
\end{aligned}
\]

Let $\cB_1$  be the trace  class equipped with the norm
$\|\cdot \|_{\cB_1}$. We recall results from \cite{K17}.

\begin{lemma}
\lb{T2.1}
 Let the potential $V$ satisfy \er{1.2}  and let $a<1$ Then   and  the operator-valued functions $Y_0(\l)$ and $Y(\l)$ are
uniformly H\"older on $\ol\C_\pm$ in the $\cB_1$--norm  and satisfy
\[
\label{e.2} \sup_{\l\in \ol\C_\pm}(1+|\l|)^{a\/2}(\|Y_0(\l)\|_{\cB_1}+\|Y(\l)\|_{\cB_1})<\iy.
\]
Moreover, they have meromorphic extensions into the whole complex plane.

\end{lemma}

\subsection{The spectral representation for $H_0$.}
We will need
some facts concerning the spectral decomposition of the Stark
operator $H_0$. Let $\cU: f\mapsto \wt f$ be the unitary transformation on $L^2(\R)$, which can be defined on $L^1(\R)\cap L^2(\R)$ by the explicit
formula
\[
\lb{UT} \wt
f(x)=(\cU f)(x)=\int_{\R} {\rm Ai}(y-x)f(y)dy,
\]
see e.g. \cite{L93},
where ${\rm Ai}(\cdot)$ is the Airy function:
\[
\label{Ai01} {\rm  Ai}(z)={1\/\pi}\int_{0}^\iy \cos
\rt({t^3\/3}+tz\rt)dt,\qqq \forall \ \ z\in \R.
\]
The unitary transformation \er{UT} carries $H_0$  over  into
multiplication by $x$ in $L^2(\R,dx)$:
\[
\label{c.3} (\cU H_0\cU^* \wt f)(x) =x\wt f(x),\qqq \wt f\in \mD(x).
\]
The Airy function Ai$(z), z\in \C$ is entire, satisfies  the equation $ {\rm Ai}''(z)=z {\rm Ai}(z)$ and the following asymptotics
\[
\label{Ai2}
\begin{aligned}
{\rm Ai}(z)={1\/2z^{1\/4}\sqrt \pi}e^{-{2\/3}z^{3\/2}}\rt(1+O(z^{-{3\/2}})\rt), \qq &
{\text if} \qq |\arg z|<\pi-\ve,\\
{\rm Ai}(-z)=\fr{1}{z^{1\/4}\sqrt \pi}\rt[ \sin \vt+O(z^{-{3\/2}}e^{|\Im
\vt|})\rt], \qq & {\text if} \qq  \ |\arg z|\le \ve, \qq
\vt={2\/3}z^{3\/2}+{\pi \/4}.
\end{aligned}
\]
as $|z|\to \iy$ uniformly in $\arg z$ for any fixed $\ve>0$ (see (4.01)-(4.05)
from \cite{O74}).

Introduce the space $L^p(\R)$ equipped by the norm
$\|f\|_p=(\int_\R|f(x)|^p dx)^{1\/p}\ge 0$ and let  $\|f\|^2=\|f\|_2^2$.  Recall results from \cite{K17}.

{\it Let  $V$ satisfy \er{1.2}. Then
the functionals $\P(\l): L^2(\R)\to \C$ given by
\[
\lb{Pf} \P(\l)f=\int_{\R} {\rm
Ai}(x-\l)|V(x)|^{1\/2}f(x)dx \qq \forall \ \l\in
\R,
\]
and the mapping $\P_1(\l)=\P(\l)^*$, for all $\l\in \R$
have analytic extensions from the real line into the whole complex plane and
satisfy
\[
\lb{Px1} \|\P(\l)\|^2 = \|\P_1(\l)\|^2 =\int_0^\g |{\rm Ai}(x-\l)|^2
|V(x)|dx \qq \forall \ \l\in \C.
\]
}

\subsection{The scattering matrix.} Recall that the S-matrix
$S(\l)$ is a scalar function of $\l\in \R$, acting as multiplication
in the fiber spaces $\C=\mH_\l$ . The stationary representation for
the scattering matrix has the form  (see e.g. \cite{RS76,J85,L93}):
\[
\label{S}
\begin{aligned}
&S(\l)=I+ \cA(\l), \qqq \l\in\R;\qqq \cA=\cA_0-\cA_1, \\
&\cA_0(\l)=-2\pi i \P(\l)V_S\P_1(\l),\qqq  \cA_1(\l)
=2\pi i \P(\l)V_SY(\l+i0)\P_1(\l),
\end{aligned}
\]
where $V_S=\sign V,\  \P_1(\l)=\P(\l)^*$. Note that due to Lemma \ref{T2.1} the operator $Y(\l\pm i0)$ is continuous in  $\l\in\R$.  The function $S(\l)$ is continuous in $\l\in \R$ and satisfies
$ S(\l)-1=O(\l^{-{1\/2}})$ as $\l\to \pm \iy$. In order to study S-matrix we define the function $X$ by
\[
\lb{dF} X(\l)=2\pi  \int_0^\g |{\rm Ai}(x-\l)|^2|V(x)|dx, \qqq \l\in \C.
\]

\begin{lemma}
\label{TA}
 Let $V$ satisfy \er{1.2}.  Then the functions $\cA_0$ and $\cA_1$ are
 continuous on the real line
 and  have  analytic extensions from
the real line into the whole upper half-plane satisfying
\[
\begin{aligned}
\label{A0i} \cA_0(\l)=-2\pi i  \int_\R {\rm Ai}(x-\l)^2 V(x)dx, \\
|\cA_0(\l)|\le X(\l),
\end{aligned}
\]
and
\[
\lb{A41} |\cA_1(\l)|\le  \|Y(\l)\| X(\l) \le  {C_a X(\l) \/ (1+|\l|)^{a\/2}},
\]
for all $\l \in \ol\C_+$ and for any fix $a<1$, where $C_a$ is some constant  depending on
$a$ and $V$.

\end{lemma}


\subsection{Fredholm determinants}

Resonances for the operator $H$ were discussed in \cite{K17}, where a central role was played by the Fredholm determinant. We recall some results from \cite{K17}.  Under condition \er{1.2} each
operator $Y_0(\l), \Im \l \ne 0,$ is trace class and thus we can
define the determinant:
\[
\label{a.2} D_\pm(\l)=\det (I +Y_0(\l)),\qqq \l\in\C_\pm.
\]
Here the function $D_\pm(\l), \l\in \C_\pm$ is
analytic in $ \C_\pm$ and satisfies
\[
\lb{DDx}
\ol D_+(\l)= D_-(\ol\l) \qqq \forall\qq \l\in \C_+,
\]
\[
\lb{aD}
 D_\pm(\l)=1+O(\l^{-{a\/2}})\qqq   as \qqq |\l| \to \infty, \qq
 \l\in\ol\C_\pm,
\]
for any fixed  $a\in (0,1)$, uniformly with respect to $\arg \l
\in [0,\pm\pi]$.
Furthermore, the determinant $D_\pm(\l), \l\in \C_\pm$ has an analytic continuation into the entire complex plane.    Moreover, for each $\l\in \R$ the S-matrix $S(\l)$ for the operators
$H_0,H$  has the form:
\[
\label{SD}
 S(\l)={D_-(\l-i0)\/D_+(\l+i0)} \qqq \forall \ \l\in \R.
\]
 Furthermore, by \er{SD}, the S-matrix
$S(\l), \l\in \R$ has an analytic extension into the whole upper
half plane $\C_+$ and a meromorphic  extension into the whole lower
half plane $\C_-$. The zeros of $S(\l),\l\in \C_+$ coincide with the
zeros of $D_-$ and the poles of $S(\l),\l\in \C_-$ are precisely
the zeros of $D_+$.

In \cite{K17} we defined the resonances of the perturbed
Stark operator $H$ as the zeros of the analytic continuation of the
determinant $D_+(\l), \l\in \C_+ $ in the lower half-plane $\C_-$.
Due to \er{DDx} the sets of zeros of $D_\pm$ in $\C_\mp$ are symmetric with respect to the real line.
Thus in order to study resonances it is enough to consider $ D_+$ or $ D_-$.
By the identity \er{SD}, the resonances equivalently can be
characterized as zeros of the scattering matrix in the lower
half plane $\C_+$ or the poles of the scattering matrix in the lower
half plane $\C_-$. Note, however, that the Riemann surface of $S(\l)$ is
the complex plane $\C$, while for the determinant the natural domain
of analyticity consists of two disconnected copies of $\C$,
corresponding to analytic continuation from $\C_+$ to $\C_-$ and
vice versa.

\subsection{Estimates on Airy functions.}

In order to study S-matrix we need the asymptotics of the Airy function
from \er{Ai2}. Furthermore we have
\[
\lb{sym1}
\begin{aligned}
\ca (x-\l)^{-{1\/4}}=(-\l)^{-{1\/4}}\big(1+{O(|\l|^{-1}}  \big),\\
(x-\l)^{{3\/2}}=(-\l)^{3\/2}+{3\/2}x(-\l)^{1\/2}+{O(|\l|^{-{1\/2}}})\ac,\qqq
|\arg \l| \geq \ve,
\end{aligned}
\]
and
\[
\lb{sym1x}
\begin{aligned}
\ca (\l-x)^{-{1\/4}}=(\l)^{-{1\/4}}\big(1+{O(|\l|^{-1}}  \big),\\
(\l-x)^{{3\/2}}=(\l)^{3\/2}-{3\/2}x(\l)^{1\/2}+{O(|\l|^{-{1\/2}}})\ac,\qqq
|\arg \l | \leq \ve,
\end{aligned}
\]
locally uniformly in $x \in \R$, as  $|\l|\to \iy$.
Here and bellow we use the following definitions
\[
\lb{dL}
\begin{aligned}
\l=|\l|e^{i\vp}\in \ol\C_+,\qq \vp\in [0, {\pi}] \qqq k=\sqrt\l\in \ol\C_+,\\
-i\l^{3\/2}=-i|\l|^{3\/2}(c+is)=|\l|^{3\/2}s-i|\l|^{3\/2}c,\ \ \ \ ,\\
c=\cos {3\vp\/2},\qqq
s=\sin {3\vp\/2}\in\ca  (0,1)&  \vp\in [0, \fr{2\pi}{3}]  \\
                    (-1,0) &  \vp\in [\fr{2\pi}{3}, \pi] \ac.
\end{aligned}
\]

These estimates give as $|\l| \to \infty $ and let $k=\sqrt\l\in \ol\C_+$:

$\bu$ Let $\vp\in [{\ve}, {\pi}]$ for some $\ve>0$. Then   one has
\[
\label{as1}
\begin{aligned}
{\rm Ai}(x-\l)^2={i\/4k \pi} e^{-i\F }
\rt(1 + {O(1)\/k}\rt),\qqq \F={4\/3}k^{3}-2xk,
\end{aligned}
\]
and, in particular,
\[
\label{as2}
|{\rm Ai}(x-\l)|^2=\fr{1}{4\pi |k|}
e^{\Im \F }\rt(1 + \fr{O(1)}{|k|} \rt).
\]

 $\bu$ Let $|\vp|\le{\ve}$ for some $\ve>0$. Then the following holds true:
\[
\label{as3}
\begin{aligned}
{\rm Ai}^2(x-\l)={1+ \sin \F\/2\pi k}+{O(e^{|\Im \F|})\/\l}.
\end{aligned}
\]
All estimates \er{as1}-\er{as3} are locally uniform in $x$
 on bounded intervals.

\begin{lemma}
\lb{TasAi}
Let  $\l=|\l|e^{i\vp}\in \ol\C_+$,  and $k=\sqrt {\l}\in \C_+$ and $\ve >0$.
Recall that $\F={4\/3}k^{3}-2xk$.

i) Define the function $\x(\l)={e^{-i{4\/3} \l^{3\/2}}\/2\sqrt \l},\ \l\in \ol\C_+$. Let $\vp\in [{\ve}, {\pi}]$.
 Then
 \[
\label{aA0}
\begin{aligned}
\cA_0(\l)=\x(\l)J_0(\l),
\\
J_0(\l)=
\int_0^\g e^{i2x k}V(x)
\rt(1 + {O(1)\/k}\rt)dx,
\end{aligned}
\]
\[
\label{aA1}
\begin{aligned}
X(\l)=|\x(\l)| J_1(\l), \qqq
\\
J_1(\l)=
\int_0^\g e^{-2x \Im k}|V(x)|
\rt(1+ \fr{O(1)}{k}\rt)dx,
\end{aligned}
\]

ii)  Let $|\vp|\le \ve$ and let $V_0=\int_0^\iy V(x)dx$ . Then
\[
\label{aA01}
\begin{aligned}
\cA_0(\l)+{iV_0\/ k}=-{i\/ k}\int_0^\g V(x)  \rt[\sin \F+{O(e^{|\Im \F|})\/k}\rt]dx.
\end{aligned}
\]

\end{lemma}
{\bf Proof.} Substituting asymptotics \er{as1}-\er{as3}
into \er{A0i} we obtain    \er{aA0}-\er{aA01}. \BBox

\section {proof of main theorems }
\setcounter{equation}{0}

We describe the Forbidden domain for resonances.

\no {\bf Proof of Theorem \ref{T1}.}
Let $\l=|\l|e^{i\vp}$, $ \vp\in [0,\pi]$ and $|\l|\to \iy$. Consider the case $\vp\in [{\ve}, {\pi}]$ for some $\ve>0$ and let $a<1$.
Using \er{A41},   \er{aA0},  \er{aA1} we obtain
 \[
\label{Axx}
\cA(\l)=\cA_0(\l)+\cA_1(\l)=\x(\l)(J_0(\l)+ J_1(\l)O(|\l|^{-{a\/2}})).
\]

i) Let $\vp\in [\fr{2\pi}{3}, {\pi}]$.  Then \er{Axx} and
the identity $|\x(\l)|=\fr{e^{-{4\/3} s |\l|^{3\/2}}}{2|\l|^{1\/2}}$ yield \er{aD1}, since we have
$
|J_0(\l)|=O(\|V\|_1)$ and $ |J_1(\l)|=O(\|V\|_1)$ and $s=\sin
 {3\vp\/2}\in (-1,0)$.

ii) Let $V$ satisfy \er{CV2} and let $ \vp\in [\fr{2\pi}{3}-\ve, \ve]$. Then $s=\sin {3\vp\/2}\in (0,1)$ and
 \[
\label{f1}
\begin{aligned}
\Im \l^{3\/2} = |\l|^{3\/2}\sin \fr{3}{2}\vp,\qqq \sin \fr{3}{2}\vp\ge c_\ve>0.
\end{aligned}
\]
%
%

From  Lemma \ref{TasAi} and \er{CV2} we obtain
 \[
\label{A0ax}
\begin{aligned}
J_0(\l)=\int_0^\g e^{i2x k}V(x)
\rt(1+ {O(1)\/k}\rt)dx=\fr{C_p+o(1)}{(-i2k)^p},\\
\end{aligned}
\]
and
\[
\lb{asFz}
 \begin{aligned}
  J_1(\l) \le C\int_0^\g e^{-2x \Im k}|V(x)|dx\le C\|V\|_1.
 \\
\end{aligned}
\]
Substituting \er{A0ax},  \er{asFz} into  \er{Axx} we obtain
$$
\cA(\l)
=
\x(\l)\rt(\fr{C_p+o(1)}{(-i2k)^p}
+ o(|k|^{-a})\rt)=\x(\l)\fr{C_p+o(1)}{(-i2k)^p},
$$
which yields $\er{aD2}$, since we can take $a>p$.
\BBox


Now we are ready to determine asymptotics of resonances.



\no {\bf Proof of Theorem \ref{T2}.}
i) Let $ \arg\l=\vp\in [\fr{2\pi}{3}-\ve,\fr{2\pi}{3}]$ for some small $\ve>0$ and $|\l|\to \iy$.
Let for shortness  $k=\sqrt \l\in \C_+$.  From \er{A41}, \er{aA0}, \er{aA1} we have
 \[
\label{r1}
\begin{aligned}
\cA(\l)=\cA_0(\l)+\cA_1(\l)=\x(\l)(J_0(\l)+O(|\l|^{-{a\/2}})),
\\
J_0(\l)=\int_0^\g e^{i2x k}V(x)
\rt(1+ {O(1)\/k}\rt)dx=\fr{C_p}{(-i2k)^p}+w_0(\l), \qq w_0(\l)=\fr{O(1)}{k^{\n}},
\end{aligned}
\]
for any fix $a<1$, where $C_p$ is defined by \er{CV2}.
This yields
$$
\cA(\l)=-\fr{iC_p{e^{-i{4\/3} k^3}}}{(-i2k)^{p+1}}(1+w(\l)),\qqq w(\l)=O(\l^{-{s\/2}}),\qquad
s=\min \{\n,a\}-p>0,
$$
for some function $w$ analytic in $\C_+$.
We have the equation for zeros of $S$:
\[
\label{r2}
\begin{aligned}
1-\fr{iC_p{e^{-i{4\/3} k^3}}}{(-i2k)^{p+1}}(1+w(\l))=0.
\end{aligned}
\]
We rewrite this equation in terms of the new variable $z={4\/3} \l^{3\/2}\in \C_+$.
From \er{r2} we obtain that the corresponding zeros satisfy the following equation:
\[
\label{r3}
\begin{aligned}
1+g(z)-\fr{e^{-iz+iz_*}}{z^{b}}=0,
\end{aligned}
\]
where $g(z):={w(\l(z))\/1+w(\l(z))}=O(z^{-{s\/3}})$ as $|z|\to\iy$ and
\[
\label{r4}
\begin{aligned}
b={p+1\/3},\qqq  2k=(3z)^{1\/3}\in \C_+,\qqq
z_*={\pi\/2}(p+2)+i\log {3^b\/C_p},\\
 (-i2k)^{p+1}=(3z)^be^{-i{\pi\/2}(p+1)}=
z^be^{-i{\pi\/2}(p+1)+b\log 3}.
\end{aligned}
\]
All zeros of the equation \er{r3} were determined in Lemma \ref{TA2} and from this lemma we have
\[
\lb{F1}
\begin{aligned}
z_n=z_n^o+z_*-b\pi+O(n^{-s/3}) \qq \as \qq n\to -\iy,
\end{aligned}
\]
where   $z_n^o$ is defined by \er{dxn}.
Then these asymptotics for give  $\l_n^-=({3\/4}z_{-n})^{2\/3}$:
$$
\begin{aligned}
\l_n^-=\m_n\rt(1-{iy_n^o+z_*-\pi b\/x_n^o}+{O(1)\/n^{{s\/3}+1}}\rt)^{2\/3}=
\m_n\rt(1-{2\/3}{iy_n^o+z_*-\pi b\/x_n^o}+{O(1)\/n^{{s\/3}+1}}\rt)
\end{aligned}
$$
where   $\m_n=({3\/4}x_{-n}^o)^{2\/3}=|x_n^o|^{2\/3}e^{i{2\/3}\pi }$,
which yields \er{arn1} for $\l_n^-, n\to \iy$.

ii) Let $\vp\in (0,\ve)$ and let $k=\sqrt \l\in \ol\C_+$ and $\F={4\/3}k^{3}-2xk$. Then
from \er{aA01} we obtain
$$
\begin{aligned}
\cA_0(\l)=-i{V_0\/k}+\cA_{01}(\l)+\cA_{02}(\l),\qqq
\cA_{02}(\l)= {1\/ |\l|}\int_0^\g |V(x)|O(e^{|\Im \F|})dx,
\\
\end{aligned}
$$
where
$$
\begin{aligned}
\cA_{01}(\l)=-{i\/ k}\int_0^\g V(x)\sin \F dx
={1\/2k}\int_0^\g V(x)  (e^{-i\F}-e^{i\F})dx
={1\/2k}\int_0^\g V(x) (e^{-i\F}+O(1))dx
\end{aligned}
$$
since for $|k|\to \iy$ we have
$$
\Im \F(\l,x)={2|k|}\rt(2|k|^2{\sin 3\f\/3}-x\sin \f\rt)\sim {2|k|\f}(2|k|^2-x)>
4\f|k|^3.
$$
Similar arguments and  \er{A41} yield for any fix $a<1$:
$$
\cA_1(\l)=\x(\l){O(1)\/k^a}.
$$
Thus collecting asymptotics of $\cA_0$ and $\cA_1$ we obtain
 \[
\label{A0g}
\begin{aligned}
\cA(\l)=\x(\l)\rt(\int_0^\g V(x)e^{i2xk} dx+ {O(1)\/k} +{O(1)\/k^a} \rt)=
\fr{ e^{-i{4\/3} k^3}C_p}{2k(-i2k)^p}\rt(1+\fr{O(1)}{k^s}\rt).
\end{aligned}
\]
This yields
$$
\cA(\l)=-\fr{iC_p{e^{-i{4\/3} k^3}}}{(-i2k)^{p+1}}(1+g(\l)),\qqq g(\l)=O(k^{-s}),\qquad
s=\min \{\n,a\}-p.
$$
We have the equation for zeros of $S$:
\[
\label{eresx}
\begin{aligned}
1-\fr{iC_p{e^{-i{4\/3} k^3}}}{(-i2k)^{p+1}}(1+g(\l))=0,
\end{aligned}
\]
We rewrite this equation in terms of the new variable $z={4\/3} \l^{3\/2}\in \C_+$.
From \er{eresx} we obtain that the corresponding zeros satisfy the following equation:
\[
\label{r3x}
\begin{aligned}
1-\fr{e^{-iz+iz_*}}{z^{b}}(1+g(\l(z)))=0, \qq \Leftrightarrow \qq
{1\/(1+g(\l(z)))}-\fr{e^{-iz+iz_*}}{z^{b}}=0
\end{aligned}
\]
where
$$
\begin{aligned}
  2k=(3z)^{1\/3}\in \C_+,\qqq
 (-i2k)^{p+1}=(3z)^be^{-i{\pi\/2}(p+1)}=
z^be^{-i{\pi\/2}(p+1)+b\log 3}.
\end{aligned}
$$
From Lemma \ref{TA2},ii) we deduce that the zeros $z_n$ of the
equation \er{r3x} have the form $ z_n=z_n^o+z_*+O(n^{-{s\/3}}) $,
which yields asymptotics of $\l_n^+=(3z_n/4)^{2\/3}$  in \er{arn1}
as $n\to \iy$. \BBox

\no {\bf Proof of Corollary \ref{T3}.}
From Theorem \ref{T2} we obtain
$$
\#\{n\in \Z: |\l_n|<r\}=\#\rt\{n\in \Z: {4\/3}|\l_n|^{3\/2}<{4\/3}r^{3\/2}\rt\}=\fr{4r^{3\/2}}{3\pi}(1+o(1))\qq \as \ r\to \iy
$$
which yields \er{Nr}.
\BBox


\section {Model equations}
\setcounter{equation}{0}

\medskip

We discuss Condition C.
\begin{lemma}
\lb{TaV}
Let  $\g>0$ and $p\in (0,1)$.
 Let   $k\in \ol \C_+$ and $|k|\to \iy$.  Then
 \[
\label{aV1}
\begin{aligned}
\int_0^\g e^{ikx}\fr{dx}{x^{1-p}}=\fr{\G(p)}{(-ik)^p}+\fr{O(1)}k,
\end{aligned}
\]
uniformly in $\arg k\in [0,\pi]$, where $\G$ is the Gamma function.
\end{lemma}

{\bf Proof.} Let $k=r\o, \o=e^{i\f}, \f \in [0, {\pi}]$.   We have
 \[
\label{aV2}
\begin{aligned}
\int_0^\g e^{ikx}\fr{dx}{x^{1-p}}=\int_0^\iy e^{ikx}\fr{dx}{x^{1-p}}-J, \qqq J=\int_\g^\iy e^{ikx}\fr{dx}{x^{1-p}},
\\
J=\fr{1}{r^{p}}\int_{r\g}^\iy e^{i\o y}\fr{dy}{y^{1-p}}=\fr{1}{i\o r^{p}}\int_{r\g}^\iy \fr{de^{i\o y}}{y^{1-p}}=\fr{1}{i\o r^{p}}\rt(-\fr{e^{i\o r\g}}{(r\g)^{1-p}}+(1-p)     \int_{r\g}^\iy \fr{de^{i\o y}}{y^{2-p}} \rt),
\end{aligned}
\]
which yields $|J|\le \fr{C}{r}$.
Moreover, using the  identity 3.381 from Gradshteyn-Ryzhyk \cite{GR07}
$$
\begin{aligned}
\int_0^\iy
e^{ikx}\fr{dx}{x^{1-p}}=\fr{\G(p)}{(-ik)^p}=e^{ip{\pi\/2}-p\log
k}\G(p), \qqq (p,k)\in (0,1)\ts\ol\C_+, \\
 (-ik)^p=e^{p\log (-ik)}=e^{-ip{\pi\/2}+p\log k},
 \qq
  \log k=\log |k|+i\arg k, \qq \arg k\in [0,\pi],
\end{aligned}
$$
we obtain \er{aV1}.
\BBox


\begin{lemma} \lb{TA2}
i) Let $(b,z_*)\in (0,1)\ts \C$ and let
$
\mD_r=\ol\C_+\sm \{|z|<\r_r\}$, where the radius $\r_r=\pi(2r-1)+\Re z_*-{b\pi\/2}$
for some integer $r>1$   large enough. Consider a function
$F(z)=\fr{e^{-i(z-z_*)}}{z^b}$, $ z\in \ol\C_+$.
 Then the function
$F-1$ in the domain $\mD_r$ have only simple zeros $z_n, \pm n\ge r$
given by
\[
\lb{F1a}
\begin{aligned}
& z_n=z_n^o+z_*^\pm +bu_n+O\rt(\fr{\log^2 n}{n^2}\rt),
  \end{aligned}
\]
where
$$
\begin{aligned}
& u_n=\fr{iy_n^o+z_*}{x_n^o}, \qq \ca z_*^+=z_*\\
                                  z_*^-=z_*-b\pi \ac,\qqq
                                  z_n^o=x_n^o+iy_n^o, \qq  \ca x_n^o=2\pi n,\qq n\in \Z,\\
  y_n^o=b\log |2\pi n|\in \R_+ \ac.
  \end{aligned}
$$

ii) Let in addition $g$ be an analytic function in $\ol\C_+$ and
satisfies $g(z)=O(z^{-\b})$ as $|z|\to \iy, z\in \ol\C_+$ uniformly in $\arg z\in [0,\pi]$ for some $\b\in (0,1)$.
Then the function $F-1-g$ in the domain $\mD_r$ for some $r>0$
  large enough have only simple zeros $z_n\in \mD_r, n\ge r$ given by
\[
\lb{Fg1}
z_n^\pm=z_{\pm n}^o+z_*^\pm+O(n^{-\b})\qqq \as \qq n\to \iy.
\]
\end{lemma}
{\bf Proof.}
i) Let $z\in \mD_r$ be a zero of $F-1$.
Then letting $z_*=x_*+iy_*\in\C$, we have
$$
1=|F(z)|=\fr{e^{(y-y_*)}}{|z|^b}\qq \Rightarrow \qq y-y_*=\log |z|^b=b\log |x|+O(1/x),
$$
as $|z|\to \iy$, since $|x|\to \iy$ in this case. Let
$$
\begin{aligned}
z=z_n^o+z_*+t,\qq  t\in \C, \qq |t|<\pi, \qq
 u=\fr{iy_n^o+z_*+t}{x_n^o}=u_n+\fr{t}{x_n^o},
\end{aligned}
$$
Consider the first case, let $z=x+iy\in\mD_r$ be a zero of $F-1$ and $x\to \iy$.
The proof for the second case $x\to -\iy$ is similar.
We have
$$
\begin{aligned}
F(z_n^o+z_*+t)=\fr{e^{-i(z_n^o+t)}}{(z_n^o+z_*+t)^b}=
\fr{e^{y_n^o}}{(x_n^o)^b}\cdot  \fr{e^{-it}}{(1+u)^b}
=  \fr{e^{-it}}{(1+u)^b}.
 \end{aligned}
$$
Thus asymptotics  $u\to 0, u_n\to 0$  yield $e^{-it}\to 1$ and $t\to 0$.
Then from the equation ${e^{-it}}={(1+u)^b}$ and $ u=u_n+\fr{t}{x_n^o}\to 0, t\to 0$ we obtain
$$
1-it(1+O(t))=1+bu(1 +O(u))\qq \Rightarrow \qq -it(1+O(t))=bu(1 +O(u)),
$$
and we get \er{F1a}, since
$$ t=-\fr{bu_n}{i+\fr{b}{x_n^o}}(1
+O(u_n))=ibu_n(1 +O(u_n)).
$$

ii)  As above the zeros of $F-1-g$ have the form
$\wt z=z_n^o+z_*+\t,\qq \t\in \C, \qq |\t|<\pi$.
Substituting this into the equation $F=1+g$ we obtain
\[
\lb{Fg1x}
\begin{aligned}
F(z_n^o+z_*+\t)=
\fr{e^{y_n^o}}{(x_n^o)^b}\cdot  \fr{e^{-i\t}}{(1+u)^b}
=  \fr{e^{-i\t}}{(1+u)^b}=1+g(z_n^o+z_*+\t),
 \end{aligned}
\]
where
$$
u=\fr{iy_n^o+z_*+\t}{x_n^o}=u_n+\fr{\t}{x_n^o}\to 0, \qqq u_n\to 0
\qq  and \qq g(z_n^o+z_*+\t)=O(n^{-\b}).
$$
 This yields $e^{-i\t}\to 1$ and $\t\to 0$. Thus from \er{Fg1x}
 we obtain
 $$
1-i\t(1+O(\t))=(1+bu(1 +O(u)))(1+O(n^{-\b})\qq \Rightarrow \qq -i\t(1+O(\t))=O(n^{-\b})
 $$
Then we get $\t=O(n^{-\b})$ which yields \er{Fg1}.
\BBox


\smallskip

\footnotesize

\no {\bf Acknowledgments.} \footnotesize Our study was supported by the RSF grant  No.
15-11-30007.

\end{document}